# Dynamical Environment in the Vicinity of Asteroids with an Application to 41 Daphne

Yu Jiang[1, 2]

1. State Key Laboratory of Astronautic Dynamics, Xi'an Satellite Control Center, Xi'an 710043, China

2. School of Aerospace Engineering, Tsinghua University, Beijing 100084, China

Y. Jiang (✉) e-mail: jiangyu_xian_china@163.com (corresponding author)

**Abstract**. We studied the dynamical environment in the vicinity of the primary of the binary asteroid. The gravitational field of the primary is calculated by the polyhedron model with observational data of the irregular shape. The equilibrium points, zero-velocity surfaces, as well as Jacobi integral have been investigated. The results show that the deviations of equilibrium points are large from the principal axes of moment of inertia. We take binary asteroid (41) Daphne and S/2008 (41) 1 for example. The distribution of topological cases of equilibrium points around (41) Daphne is different from other asteroids. The topological cases of the outer equilibrium points E1-E4 are Case 2, Case 5, Case 2, and Case 1. The topological case of the inner equilibrium point E5 is Case 1. Among the four outer equilibrium points E1-E4, E4 is linearly stable and other outer equilibrium points are unstable. Considering the shape variety of the body from Daphne to a sphere, we calculated the zero-velocity surfaces and the locations as well as eigenvalues of equilibrium points. It is found that the topological case of the outer equilibrium point E2 change from Case 5 to Case 1, and its stability change from unstable to linearly stable. Using the gravitational force acceleration calculated by the polyhedron model with the irregular shape, we simulated the orbit for the moonlet in the potential of (41) Daphne.

## 1 Introductions

Since the detection of a moonlet around the asteroid 243 Ida [1] by the spacecraft Galileo in 1993, several binary asteroid systems have been found in Solar system. This make the study of dynamical environment in the gravitational potential of asteroids become significative and important, and can help to understand the dynamical behaviors of moonlets relative to the primary of the binary asteroid systems [2].

41 Daphne is a large asteroid from the main belt [3, 4]. Matter et al. [3] applied the convex 3-D shape model with the data from lightcurves and images and derived the volume equivalent diameter of asteroid (41) Daphne between 194 and 209 km. Conrad et al. [4] estimated the size of 41 Daphne to be 239×183×153 km with observation data. 41 Daphne has a moonlet S/2008 (41) 1 with the size smaller than 2 km [4]. The diameter ratio [3, 4] for the moonlet and the primary in the binary system (41) Daphne-S/2008 (41) 1 is only 0.011. The binary system (41) Daphne-S/2008 (41) 1 has the most extreme size ratio among known binary asteroid systems in the Solar system [3-6]. Thus we investigate the dynamics in the gravitational potential of (41) Daphne; the results are also useful to understand the dynamical behaviors in other large-size-ratio binary asteroid systems.

The polyhedron model can deal with the gravitational potential generated by an irregular-shaped asteroid with constant density [7-10]. Previous studies used the polyhedron model to analyze the dynamical environments around asteroids 21 Lutetia [11], 216 Kleopatra [12, 13], 433 Eros [14, 15], etc. Wang et al. [16] used the irregular shape model of asteroids generated by observed data and calculated the positions and topological cases of several contact binary asteroids, including 1996 HW1, 4769 Castalia, 25143 Itokawa, etc. Bosanac et al. [17] used the restricted three-body problem (CR3BP) to model the gravitational force of the primary in the large mass ratio binary system, and investigated the stability of motion of a massless moonlet in the potential of the primary. Our interest is to analyze the dynamical environments around the primary of large-size-ratio binary asteroid systems with considering the gravitational potential generated by the irregular shape of the primary.

This paper is split into the following sections. Section 2 deals with the polyhedron model, zero-velocity curves, and equilibrium points for 41 Daphne, and calculated eigenvalues, Jacobi integral, Hessian matrix, and topological cases of equilibrium points for 41 Daphne. Section 3 focuses on the variety of topological classification, stability, and index of inertia of the equilibrium points of the body during the shape variety from 41 Daphne to a sphere. Section 4 covers the simulation of orbits of the moonlet in the gravitational potential of 41 Daphne. The gravitational potential of 41 Daphne is computed using the polyhedron model with observation data. Finally, the Conclusion section presents a brief review of the results.

## 2 Daphne Dynamical Environment: Numerical Study

Let $\boldsymbol{\omega}$ be the rotation velocity of the asteroid, and the unit vector $\mathbf{e_z}$ be the z axis in the body-fixed frame. $\mathbf{e_z}$ is defined by $\boldsymbol{\omega} = \omega \mathbf{e_z}$. Denote $U\left(x,y,z\right)$ as the gravitational potential of the asteroid. The body-fixed frame is defined as the coordinate system of principal axis of inertia. The z, y, and x are principal axes of largest, intermediate, and smallest moment of inertia.

The gravitational potential and force [7] of the body can be calculated using the polyhedron model:

$$\begin{cases} U = -\frac{1}{2}\mathrm{G}\sigma \sum\limits_{e\in edges} \mathbf{r}_e \cdot \mathbf{E}_e \cdot \mathbf{r}_e \cdot L_e + \frac{1}{2}\mathrm{G}\sigma \sum\limits_{f\in faces} \mathbf{r}_f \cdot \mathbf{F}_f \cdot \mathbf{r}_f \cdot \omega_f \\ \nabla U = G\sigma \sum\limits_{e\in edges} \mathbf{E}_e \cdot \mathbf{r}_e \cdot L_e - G\sigma \sum\limits_{f\in faces} \mathbf{F}_f \cdot \mathbf{r}_f \cdot \omega_f \end{cases} . \quad (1)$$

Where $\nabla U$ is the gravitational force, G = 6.67 × $10^{-11}$ $m^3kg^{-1}s^{-2}$ is the Newtonian gravitational constant, $\sigma$ represents the density of the asteroid; $\mathbf{r}_e$ and $\mathbf{r}_f$ are body-fixed vectors from points to some points on the edge $e$ and face $f$ of the polyhedron, respectively; $\mathbf{E}_e$ and $\mathbf{F}_f$ are face and edge tensors, respectively; $L_e$ represents the integration factor of the point while $\omega_f$ represent the signed solid angle.

The dynamical equations around an asteroid can be expressed in the body-fixed frame as

$$\begin{cases} \ddot{x} - \dot{\omega} y - 2\omega\dot{y} - \omega^2 x + \dfrac{\partial U}{\partial x} = 0 \\ \ddot{y} + \dot{\omega} x + 2\omega\dot{x} - \omega^2 y + \dfrac{\partial U}{\partial y} = 0 \\ \ddot{z} + \dfrac{\partial U}{\partial z} = 0 \end{cases}, \tag{2}$$

where $\omega$ is the rotation speed of the asteroid, which is the norm of $\boldsymbol{\omega}$.

The effective potential is defined as

$$V = U - \frac{\omega^2}{2}\left(x^2 + y^2\right); \tag{3}$$

The Jacobi integral is

$$H = U + \frac{1}{2}\left(\dot{x}^2 + \dot{y}^2 + \dot{z}^2\right) - \frac{\omega^2}{2}\left(x^2 + y^2\right) \ . \tag{4}$$

The Jacobi integral is a constant if $\omega$ is time invariant. The physical meaning of the Jacobi integral is the relative energy of the particle.

Denote $\left(x_L, y_L, z_L\right)^T$ as an equilibrium point, let $\left(x, y, z\right)^T$ be a point near the equilibrium point. Let

$$\begin{matrix} \xi = x - x_{\mathrm{L}} \\ \eta = y - y_{\mathrm{L}} \\ \zeta = z - z_{\mathrm{L}} \end{matrix}, \quad \begin{matrix} V_{\mathrm{xx}} = \left( \dfrac{\partial^2 V}{\partial x^2} \right)_{\mathrm{L}} \\ V_{\mathrm{yy}} = \left( \dfrac{\partial^2 V}{\partial y^2} \right)_{\mathrm{L}} \\ V_{\mathrm{zz}} = \left( \dfrac{\partial^2 V}{\partial z^2} \right)_{\mathrm{L}} \end{matrix} \quad \text{and} \quad \begin{matrix} V_{\mathrm{xy}} = \left( \dfrac{\partial^2 V}{\partial x \partial y} \right)_{\mathrm{L}} \\ V_{\mathrm{yz}} = \left( \dfrac{\partial^2 V}{\partial y \partial z} \right)_{\mathrm{L}} \\ V_{\mathrm{xz}} = \left( \dfrac{\partial^2 V}{\partial x \partial z} \right)_{\mathrm{L}} \end{matrix} \tag{5}$$

Then the linearised equations [18] of motion for a massless particle relative to the equilibrium point can be written as

$$\mathbf{M}\ddot{\mathbf{X}} + \mathbf{G}\dot{\mathbf{X}} + \mathbf{KX} = 0, \tag{6}$$

where

$$\mathbf{X} = \begin{bmatrix} \xi & \eta & \zeta \end{bmatrix}^{\mathrm{T}}, \ \mathbf{M} = \begin{pmatrix} 1 & 0 & 0 \\ 0 & 1 & 0 \\ 0 & 0 & 1 \end{pmatrix}, \ \mathbf{G} = \begin{pmatrix} 0 & -2\omega & 0 \\ 2\omega & 0 & 0 \\ 0 & 0 & 0 \end{pmatrix}, \ \mathbf{K} = \begin{pmatrix} V_{\mathrm{xx}} & V_{\mathrm{xy}} & V_{\mathrm{xz}} \\ V_{\mathrm{xy}} & V_{\mathrm{yy}} & V_{\mathrm{yz}} \\ V_{\mathrm{xz}} & V_{\mathrm{yz}} & V_{\mathrm{zz}} \end{pmatrix}.$$

The characteristic equation of the above equation satisfies

$$\begin{vmatrix} \lambda^2 + V_{\mathrm{xx}} & -2\omega\lambda + V_{\mathrm{xy}} & V_{\mathrm{xz}} \\ 2\omega\lambda + V_{\mathrm{xy}} & \lambda^2 + V_{\mathrm{yy}} & V_{\mathrm{yz}} \\ V_{\mathrm{xz}} & V_{\mathrm{yz}} & \lambda^2 + V_{\mathrm{zz}} \end{vmatrix} = 0, \tag{7}$$

where $\lambda$ represents the eigenvalues of the equilibrium point.

In Figure 1, we plot the zero-velocity curves and equilibrium points around asteroid 41 Daphne. The zero-velocity curves and equilibrium points are showed in the equatorial plane of Daphne. From Figure 1, one can see that there are five equilibrium points in the potential of Daphne. Figure 2 shows projection of zero-velocity curves in different planes.

To show the values of positions and Jacobi integrals of equilibrium points numerically, we also calculated these values and presented the relative quantity to help to analyze. Table 1 presents positions and Jacobi integrals of equilibrium points around 41 Daphne, while Table 2 presents relative distance of the equilibrium point position and coordinate axis. In Table 2, x axis equilibrium points represent equilibrium points which are near x axis; the same for y axis equilibrium points.

Several previous literatures used symmetric simple-shaped bodies to model the gravitational field of asteroids. These symmetric simple-shaped bodies include cube model [19], dumbbell-shaped model [20], ellipsoid-sphere model [21, 22], etc. Feng et al. [21] used the ellipsoid-sphere body to model the gravitational field of contact binary asteroid 1996 HW1, and calculated the locations of outside equilibrium points around 1996 HW1. A contact binary asteroid is also a single asteroid, the shape of a contact binary asteroid consists of two parts, and the two parts are connected by a neck. Liang et al. [22] also used the ellipsoid-sphere body to model the gravitational field of 1996 HW1, they also presents the locations of outside equilibrium points around 1996 HW1. Using the ellipsoid-sphere body to calculate, the locations of equilibrium points around asteroid 1996 HW1 are symmetric relative to one axis of the coordinate system; two equilibrium points are on the principal axes of largest moment of inertia; in addition, eigenvalues of the other two equilibrium points are the same. But in fact, the locations of equilibrium points around asteroids are not symmetric; the locations of equilibrium points around asteroids are not on principal axes of largest, intermediate, or smallest moment of inertia; different equilibrium points have different eigenvalues. From Table 2, one can see that the locations of equilibrium points around asteroid 41 Daphne are non-symmetric. The x axis equilibrium point E1 deviate x axis, it has significantly components on y axis and z axis. The relative distance of the equilibrium point E1 and coordinate axis y and z are 12.0115% and 2.29486%, respectively. For the x axis equilibrium point E3, the relative distance of the equilibrium point and coordinate axis y and z are 34.2296% and 2.59659%, respectively. The deviations of equilibrium points are so large and cannot be neglected. For the y axis equilibrium points E2 and E4, the deviations of equilibrium points are also large. The relative distance of the equilibrium point E2 and coordinate axis x and z are 9.64194% and 0.873343%, respectively. For the equilibrium point E4, these two values are 27.2740% and 0.762346%, respectively. To see the asymmetry of the equilibrium points, we also plot Figure 3 to show locations of equilibrium points around asteroid 41 Daphne viewed in different planes.

The distribution of eigenvalues of the equilibrium point can confirm the topological cases and stability of the equilibrium point. Table 3 gives eigenvalues of these equilibrium points. Table 4 gives the topological classification, stability, and index of inertia of equilibrium points of asteroid 41 Daphne. Table A1 in Appendix A presents Hessian matrix of the equilibrium points around Daphne. The positive/negative index of inertia for an equilibrium point is defined as the positive/negative index of inertia of the Hessian matrix of the equilibrium points. The topological cases [16, 18] are defined by distributions of eigenvalues, i.e. Case 1: $\pm i\beta_j\left(\beta_j \in \mathrm{R}, \beta_j > 0; j = 1,2,3\right)$; Case 2: $\pm\alpha_j\left(\alpha_j \in \mathrm{R}, \alpha_j > 0, j = 1\right)$, $\pm i\beta_j\left(\beta_j \in \mathrm{R}, \beta_j > 0; j = 1,2\right)$; Case 5: $\pm i\beta_j\left(\beta_j \in \mathrm{R}, \beta_j > 0, j = 1\right)$, $\pm\sigma \pm i\tau\left(\sigma, \tau \in \mathrm{R}; \sigma, \tau > 0\right)$. From Table 3, one can see that eigenvalues of equilibrium points E1 and E3 are different, and eigenvalues of equilibrium points E2 and E4 are also different. The forms of eigenvalues of equilibrium point E2 are $\pm i\beta_j \left( \beta_j \in \mathrm{R}, \beta_j > 0; j = 1 \right)$ and $\pm\sigma \pm i\tau \left( \sigma, \tau \in \mathrm{R}; \sigma, \tau > 0 \right)$ , while the forms of eigenvalues of equilibrium point E4 are $\pm i\beta_j \left( \beta_j \in \mathrm{R}, \beta_j > 0; j = 1, 2, 3 \right)$; in other words, the topological cases of equilibrium points E2 and E4 are Case 5 and Case 1, respectively. Equilibrium point E2 is unstable while E4 is linearly stable.

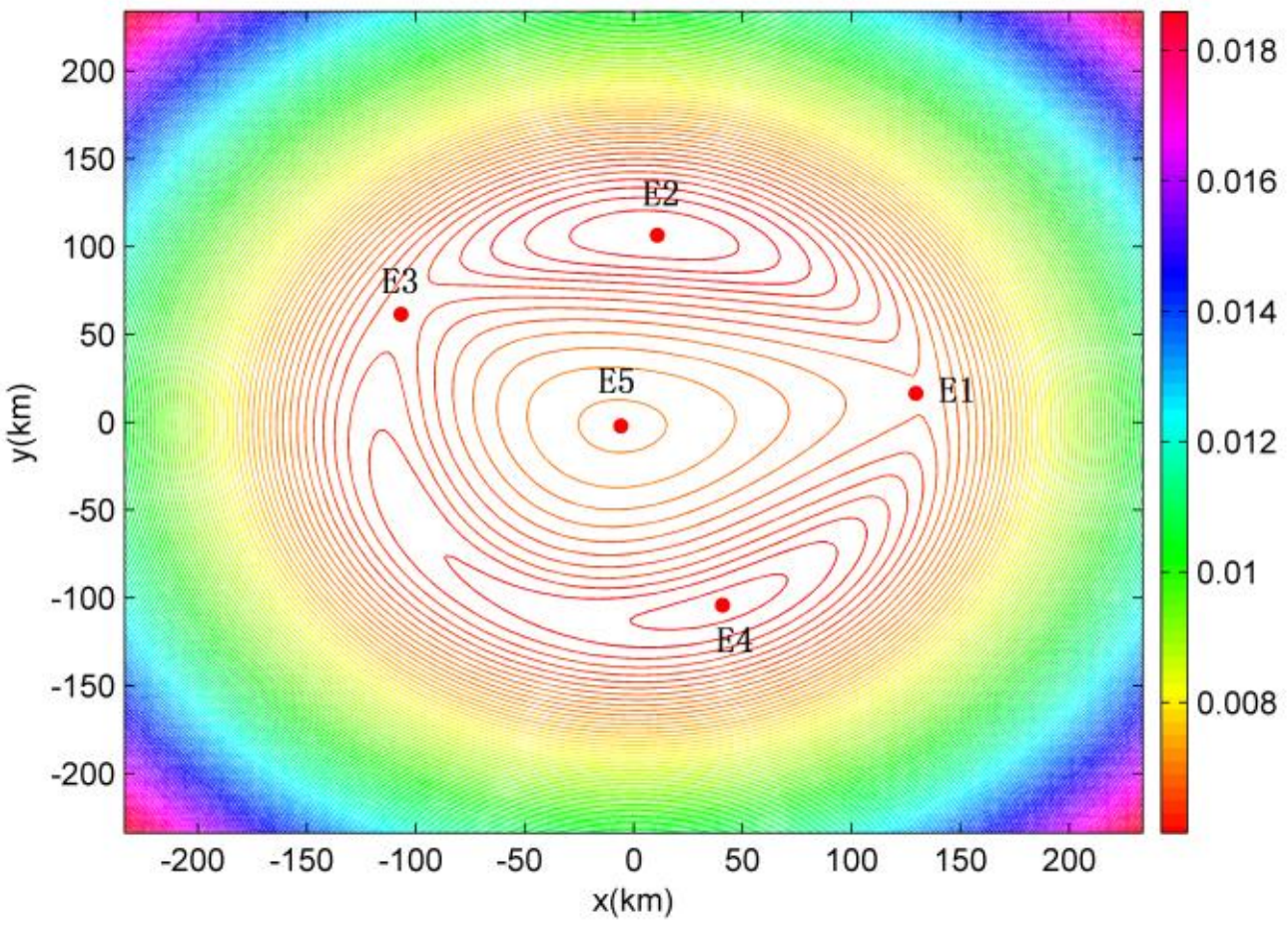

(a)

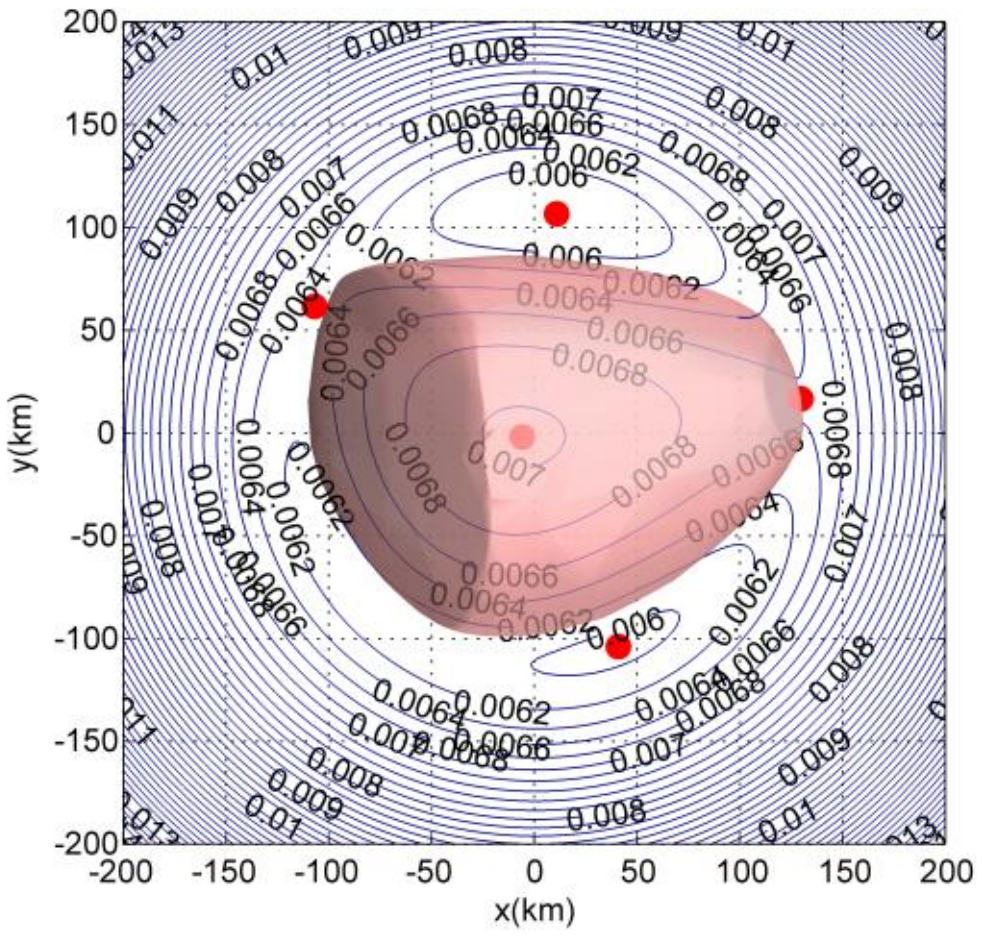

(b)

Figure 1. Zero-velocity curves and equilibrium points around asteroid 41 Daphne. (a) Projections of zero-velocity curves and equilibrium points in the equatorial plane; (b) Contour plot of zero-velocity curves and equilibrium points relative to the body of asteroid 41 Daphne.

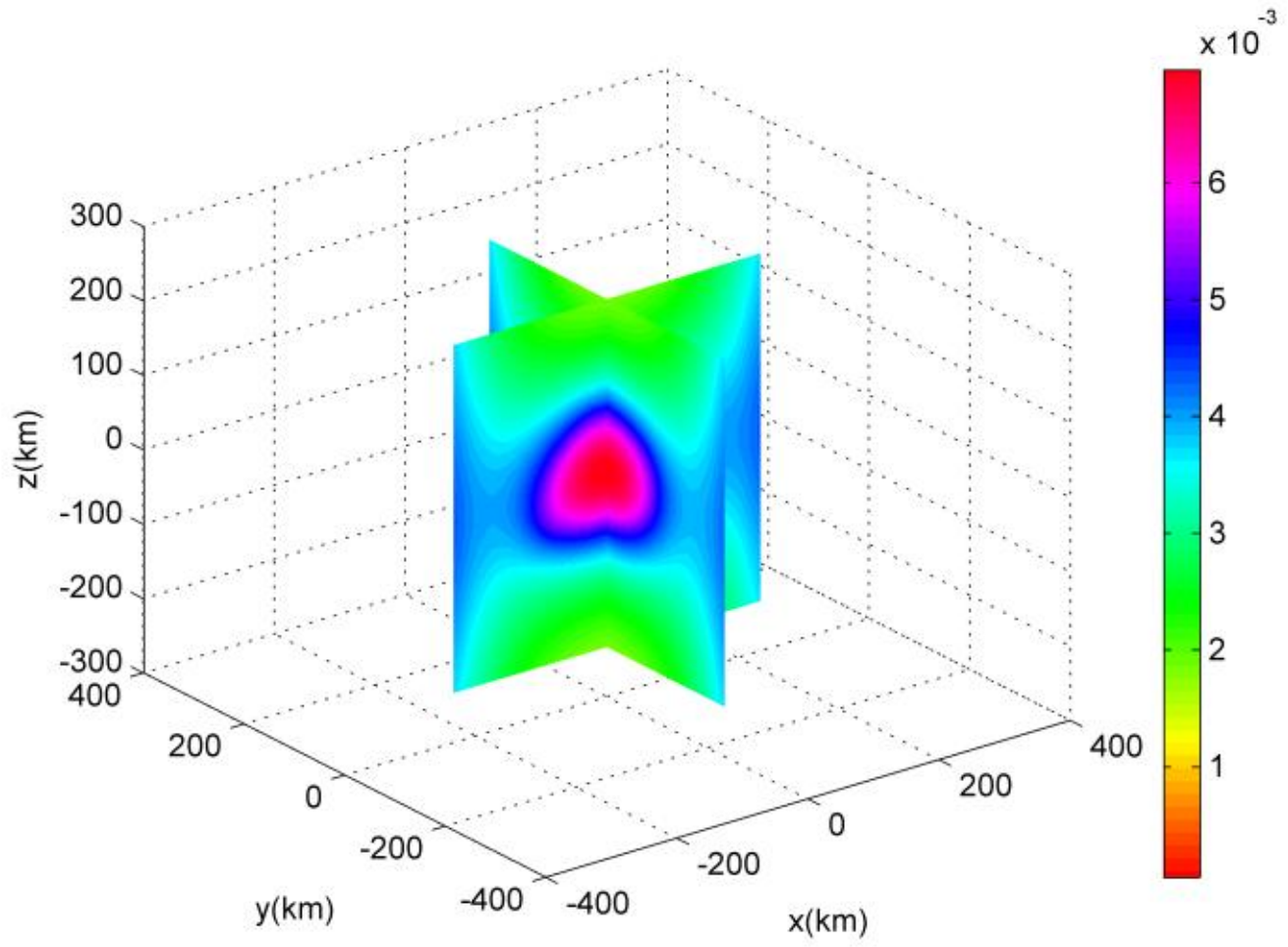

(a)

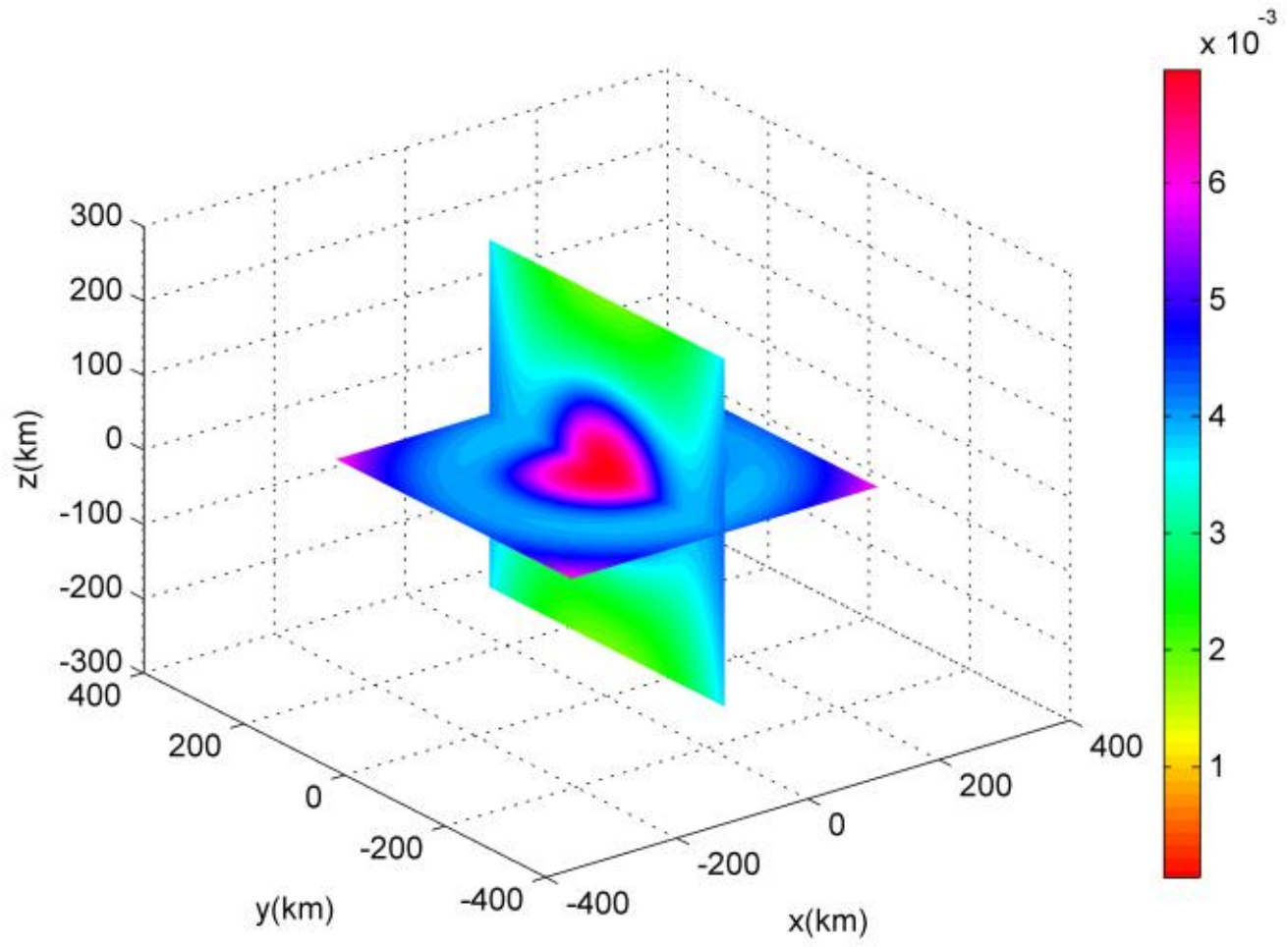

(b)

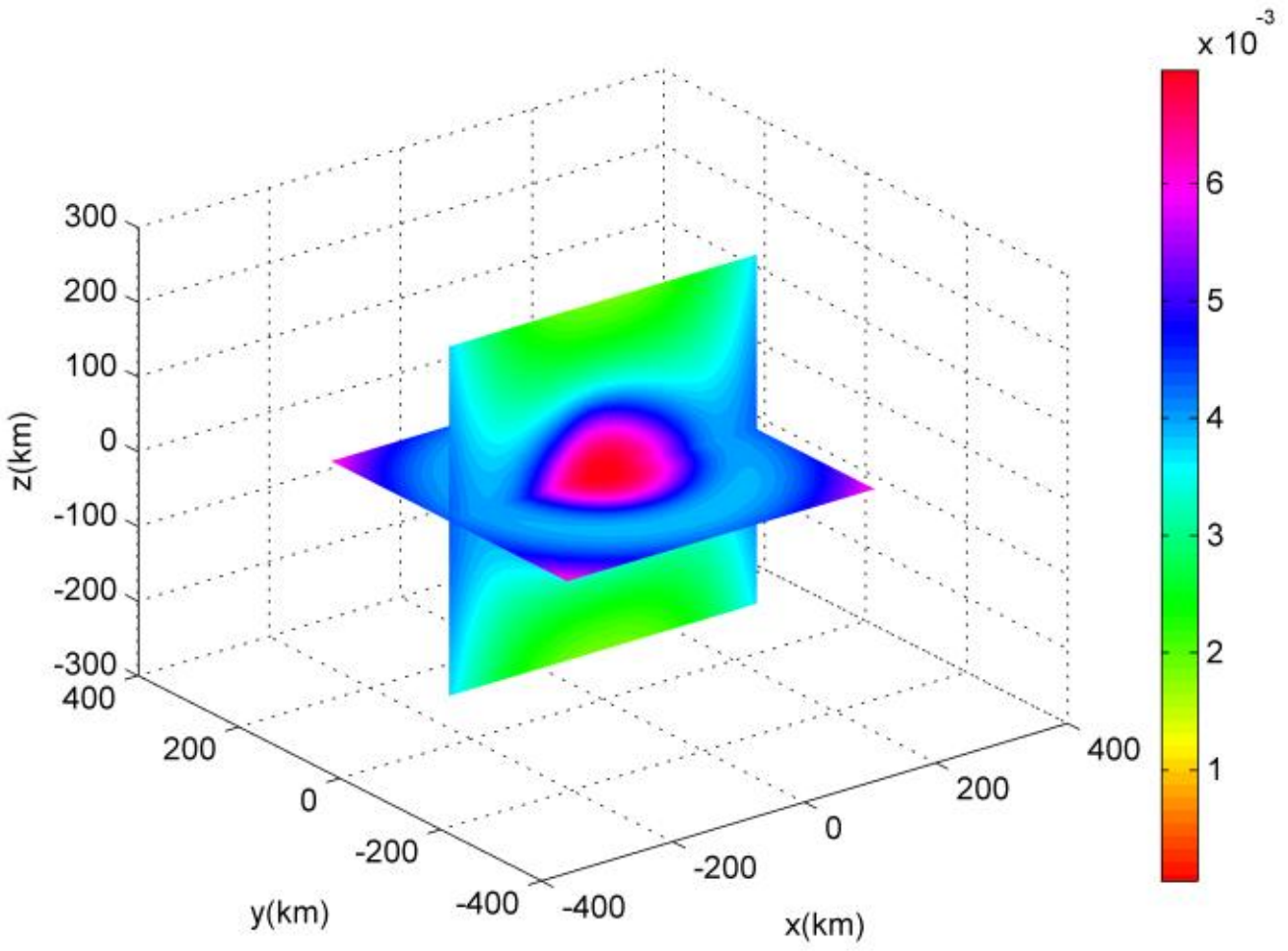

(c)

Figure 2. Projection of zero-velocity curves in different planes. (a) viewed in the xz and yz planes; (b) viewed in the xy and yz planes; (c) viewed in the xy and xz plane.

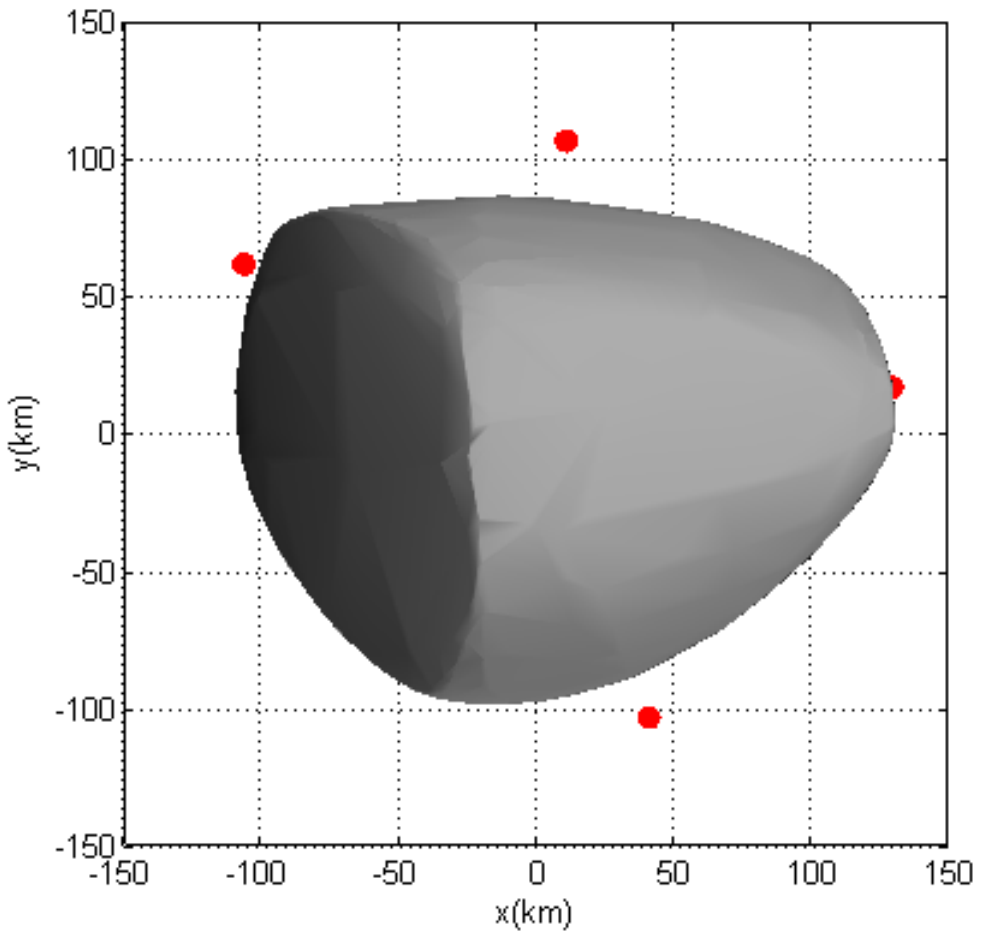

(a)

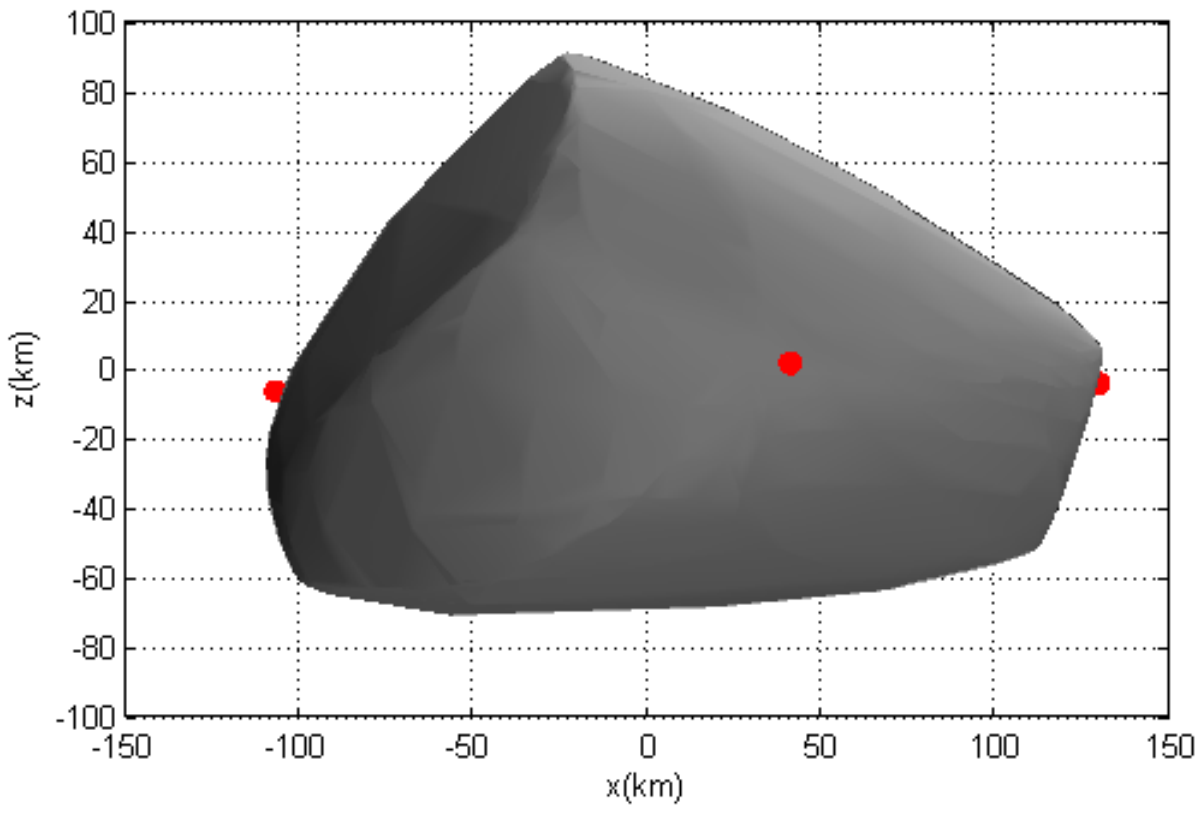

(b)

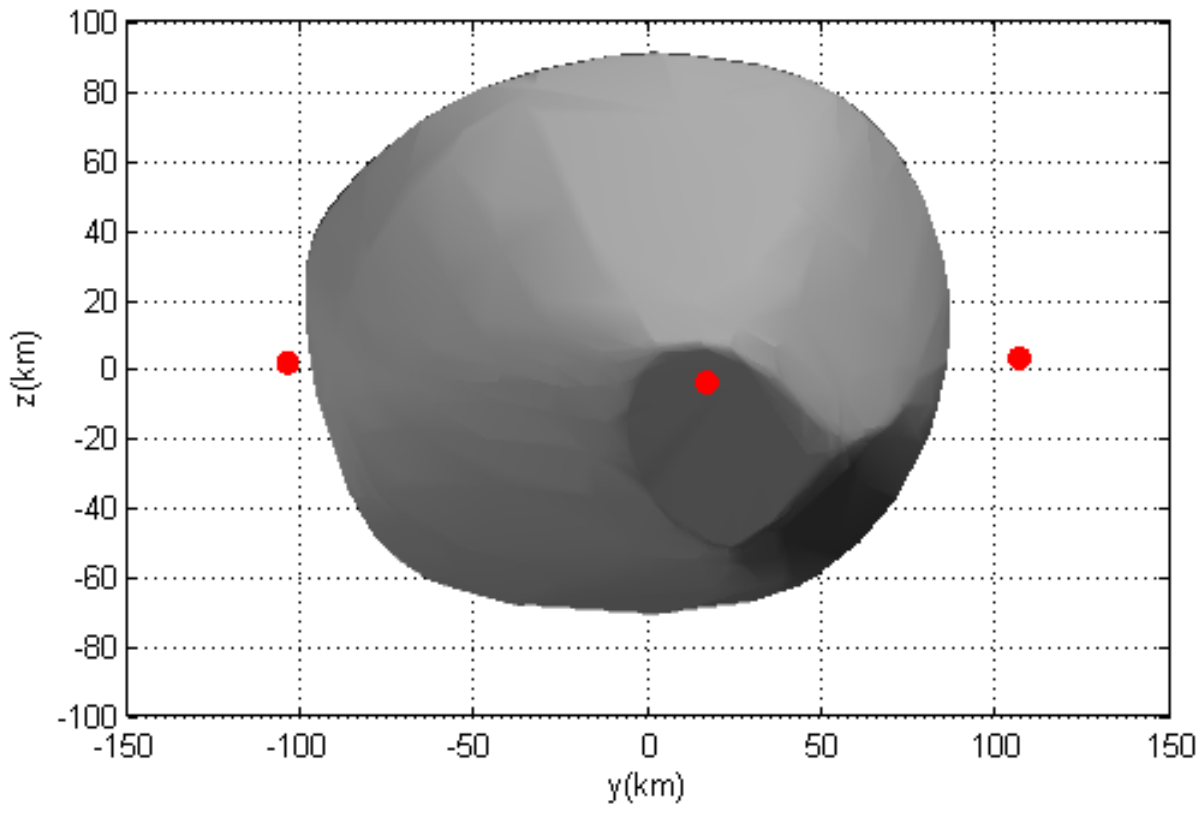

(c)

Figure 3. Locations of equilibrium points around asteroid 41 Daphne. (a) viewed in the xy plane; (b) viewed in the xz plane; (c) viewed in the yz plane.

Table 1 Positions and Jacobi integrals of the equilibrium points around 41 Daphne

| Equilibrium Points | x (km) | y (km) | z (km) | Jacobi integral ($1.0\times 10^{-3}km^2s^{-2}$) |
|---|---|---|---|---|
| E1 | 181.315 | 21.7786 | -4.16092 | 4.56987 |
| E2 | 16.2431 | 168.463 | 1.47126 | 4.42396 |
| E3 | -168.611 | 57.7148 | -4.37813 | 4.52802 |
| E4 | 45.1892 | -165.686 | 1.26310 | 4.44930 |
| E5 | 1.99899 | -0.989692 | 1.33615 | 8.09998 |

Table 2 Relative distance of the equilibrium point and coordinate axis

x axis equilibrium points

| Equilibrium Points | $\left\|\frac{y}{x}\right\| \times 100\%$ | $\left\|\frac{z}{x}\right\| \times 100\%$ |
|---|---|---|
| E1 | 12.0115% | 2.29486% |
| E3 | 34.2296% | 2.59659% |

y axis equilibrium points

| Equilibrium Points | $\left\|\frac{x}{y}\right\| \times 100\%$ | $\left\|\frac{z}{y}\right\| \times 100\%$ |
|---|---|---|
| E2 | 9.64194% | 0.873343% |
| E4 | 27.2740% | 0.762346% |

Table 3 Eigenvalues of the equilibrium points around 41 Daphne

| Equilibrium Points(× $10^{-3}s^{-1}$) | $\lambda_1$ | $\lambda_2$ | $\lambda_3$ | $\lambda_4$ | $\lambda_5$ | $\lambda_6$ |
|---|---|---|---|---|---|---|
| E1 | 0.339730i | -0.339730i | 0.328400i | -0.328400i | 0.219650 | -0.219650 |
| E2 | 0.301601i | -0.301601i | 0.087732+ 0.222988i | 0.087732- 0.222988i | -0.087732+ 0.222988i | -0.087732- 0.222988i |
| E3 | 0.326157i | -0.326157i | 0.301477i | -0.301477i | 0.149163 | -0.149163 |
| E4 | 0.306565i | -0.306565i | 0.203596i | -0.203596i | 0.198957i | -0.198957i |
| E5 | 0.992021i | -0.992021i | 0.818859i | -0.818859i | 0.394811i | -0.394811i |

Table 4. The topological classification, stability, and index of inertia of the equilibrium points of asteroid 41 Daphne. LS: linearly stable; U: unstable; P: positive definite; N: non-positive definite; Index of inertia: positive/ negative index of inertia

| Equilibrium Points | Topological Case | Stability | $\nabla^2 V$ | Index of Inertia |
|---|---|---|---|---|
| E1 | 2 | U | N | 2/1 |
| E2 | 5 | U | N | 1/2 |
| E3 | 2 | U | N | 2/1 |
| E4 | 1 | LS | N | 1/2 |
| E5 | 1 | LS | P | 3/0 |

## 3 Dynamical Environment with the Shape Change of Daphne

Considering asteroid 41 Daphne has a highly irregular shape, and the deviations of equilibrium points relative to the coordinate axes of the principal inertia frame are large. We change the shape of this asteroid to see the variety of zero-velocity curves and equilibrium points.

We use the homotopy method to generate the shape variety of the body. The new shape of the body is calculated by

$$S_{new} = \eta K + \left(1-\eta\right) S , \tag{8}$$

where $K$ represents the shape of asteroid 41 Daphne, $S$ represents the shape of a sphere, $\eta$ changes from 0 to 1. The radius of the sphere is set to be 130.71 km. When $\eta$ equal 0, the new shape is a sphere; while $\eta$ equal 1, the new shape and the shape of asteroid 41 Daphne are the same. Figure 4 shows zero-velocity curves and equilibrium points with the shape variety of the body.

When $\eta$=1.0, the shape has no change relative to the shape of Daphne, Table 3 gives the eigenvalues of the equilibrium points while Table 4 gives the topological classification, stability, and index of inertia of the equilibrium points.

In Figure 4, we present four cases, i.e. $\eta$=0.8, 0.6, 0.4, and 0.2. Table A2 in Appendix A presents positions and Jacobi integrals of the equilibrium points of the body during the shape variety. From Figure 4 and Table A2, one can see that although the locations of equilibrium points have obvious change, the deviations of outside equilibrium points relative to the principal axis of the moment of inertia keep large during the shape variety from Daphne to a sphere. The outside equilibrium points include E1, E2, E3, and E4. The deviations of inner equilibrium points E5 relative to the principal axis of the moment of inertia decrease during the shape variety from Daphne to a sphere.

Eigenvalues of the equilibrium points of the body during the shape variety have been presented in Table A3 in Appendix A. Comparing Table A3 with Table 3, one can see that the topological cases of equilibrium points with the parameter $\eta$=0.8 and the topological cases of equilibrium points with the parameter $\eta$=1.0 are the same. However, when $\eta$=0.6, the topological case of equilibrium point E2 varies from Case 5 to Case 1, and the form of eigenvalues of equilibrium point E2 change from $\begin{cases} \pm i\beta_j \left(\beta_j \in \mathrm{R}, \beta_j > 0, j = 1\right) \\ \pm\sigma \pm i\tau\left(\sigma, \tau \in \mathrm{R}; \sigma, \tau > 0\right) \end{cases}$ to $\pm i\beta_j \left(\beta_j \in \mathrm{R}, \beta_j > 0; j = 1, 2, 3\right)$. The forms of eigenvalues of equilibrium points for $\eta$=0.6, 0.4, and 0.2 are the same. Table A4 in Appendix A gives the topological classification, stability, and index of inertia of the equilibrium points of the body during the shape variety. From Table A2 and A3, one can see that the locations, Jacobi integrals, as well as eigenvalues of equilibrium points vary during the variety of the shape of the body. During the variety of the shape of the body, i.e, $\eta$ changes from 1.0 to 0.2, locations of outside equilibrium points E1-E4 move far away from the center of the body, the norm of position vectors of the equilibrium points E1-E4 magnify. However, the location of inner equilibrium point E5 moves close to the center of the body, the norm of the position vector of the equilibrium point E5 decrease. The values of Jacobi integrals of equilibrium points increase during variety of the shape from Daphne to a sphere. The Jacobi integral of equilibrium point E2 is the smallest, while the Jacobi integral of equilibrium point E5 is the biggest. From Table A4, one can see that the topological classification and stability of equilibrium points E2 vary during the variety of the shape of the body, and the Hessian matrix of effective potential and the index of inertia remain unchanged.

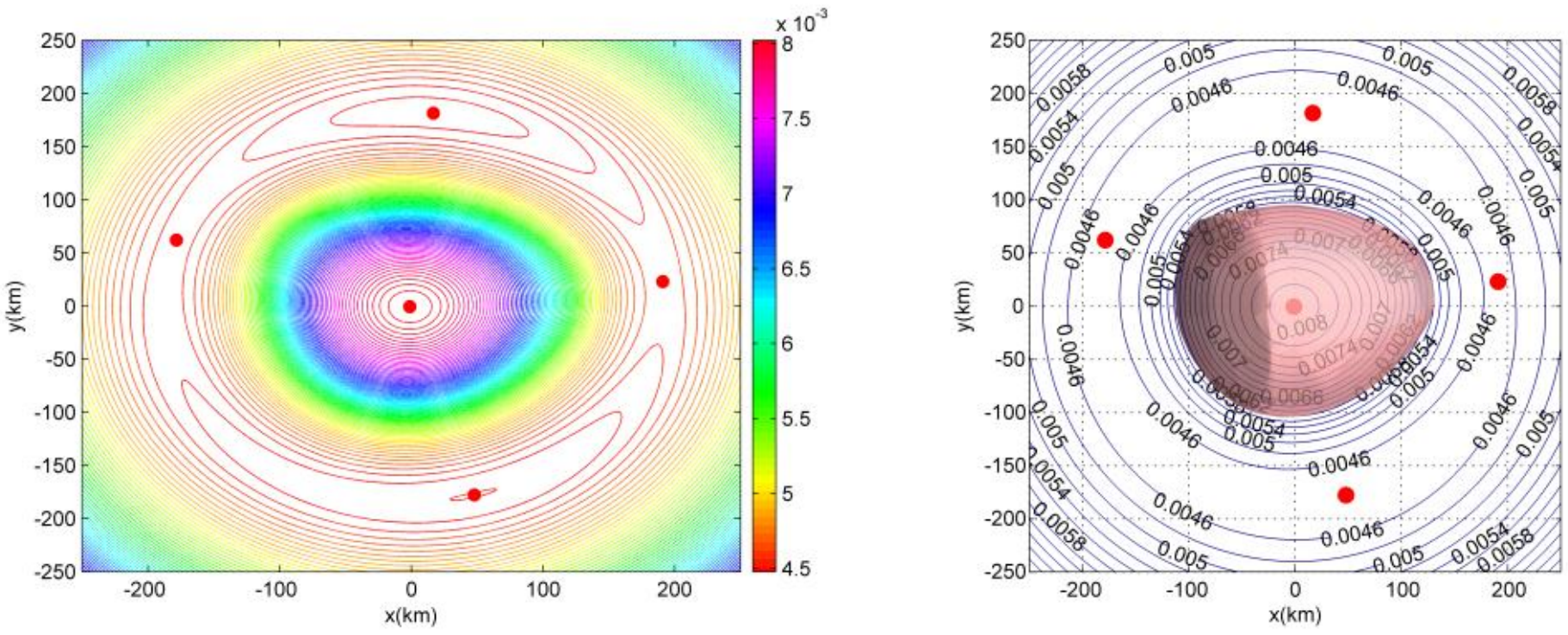

(a)

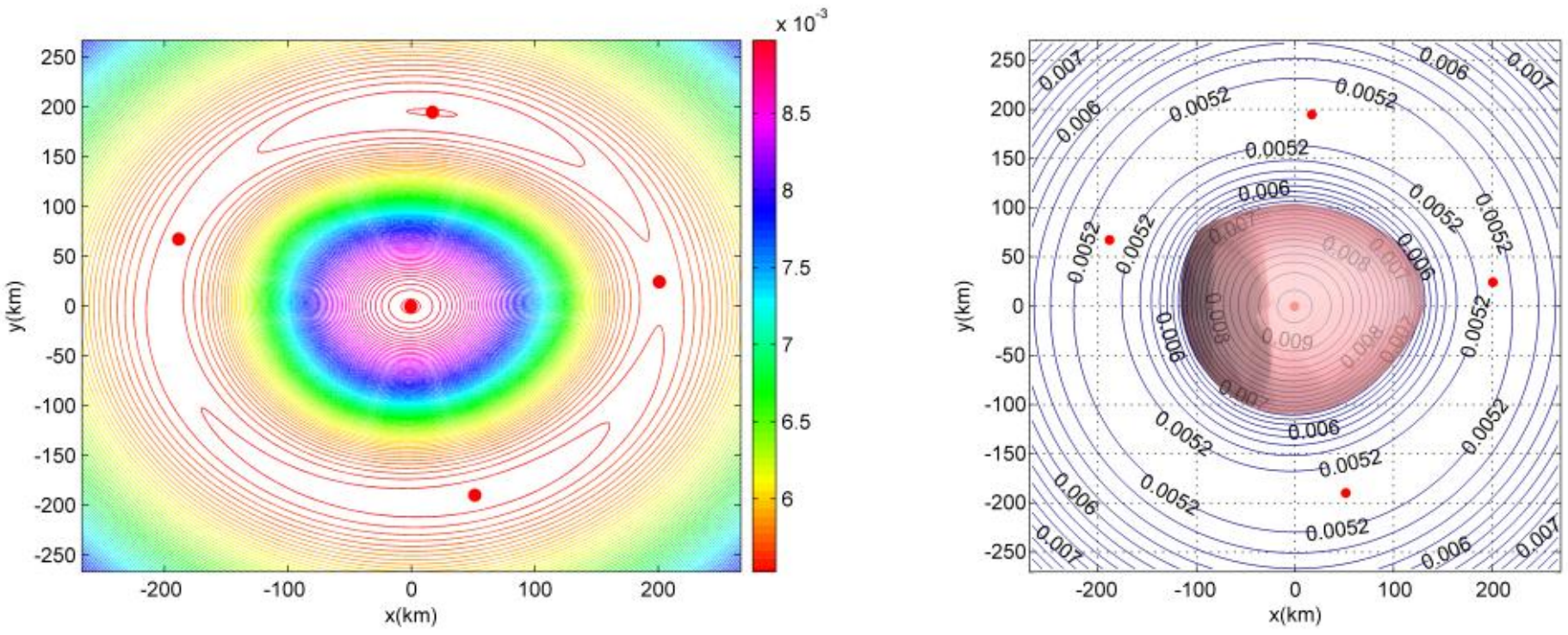

(b)

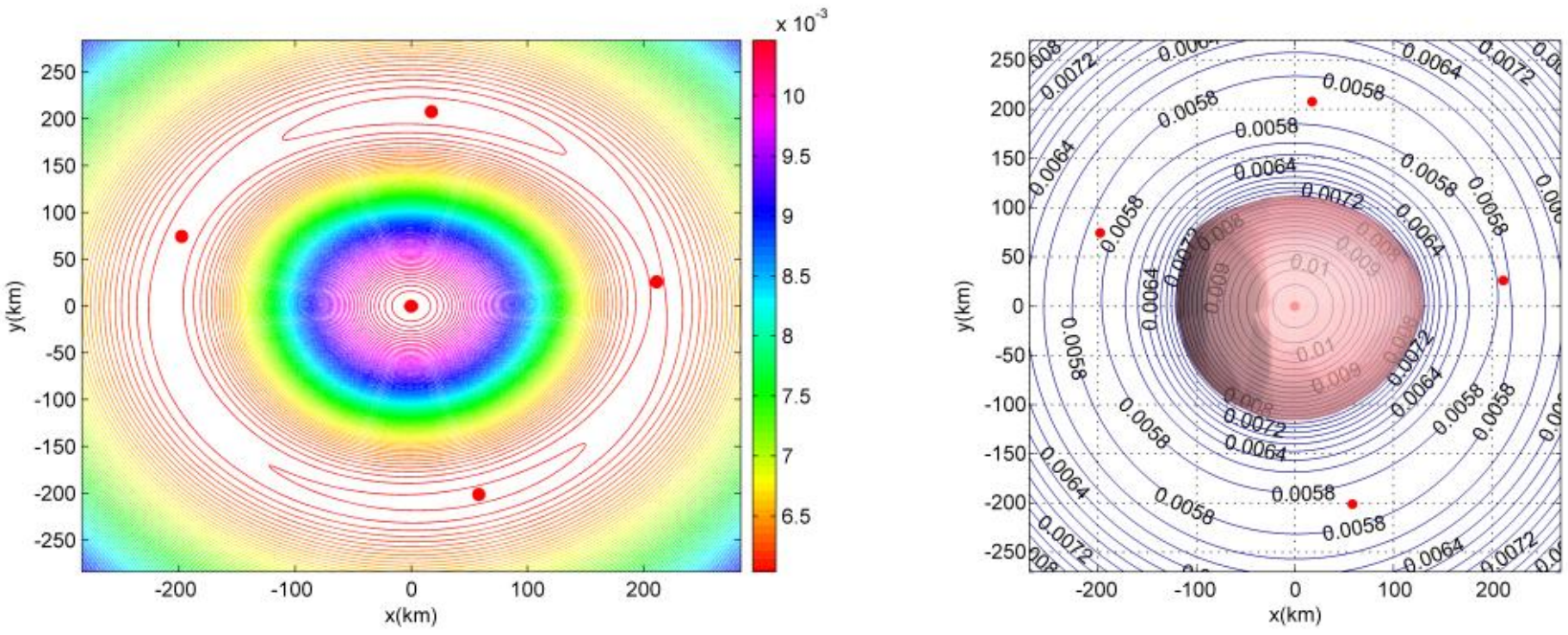

(c)

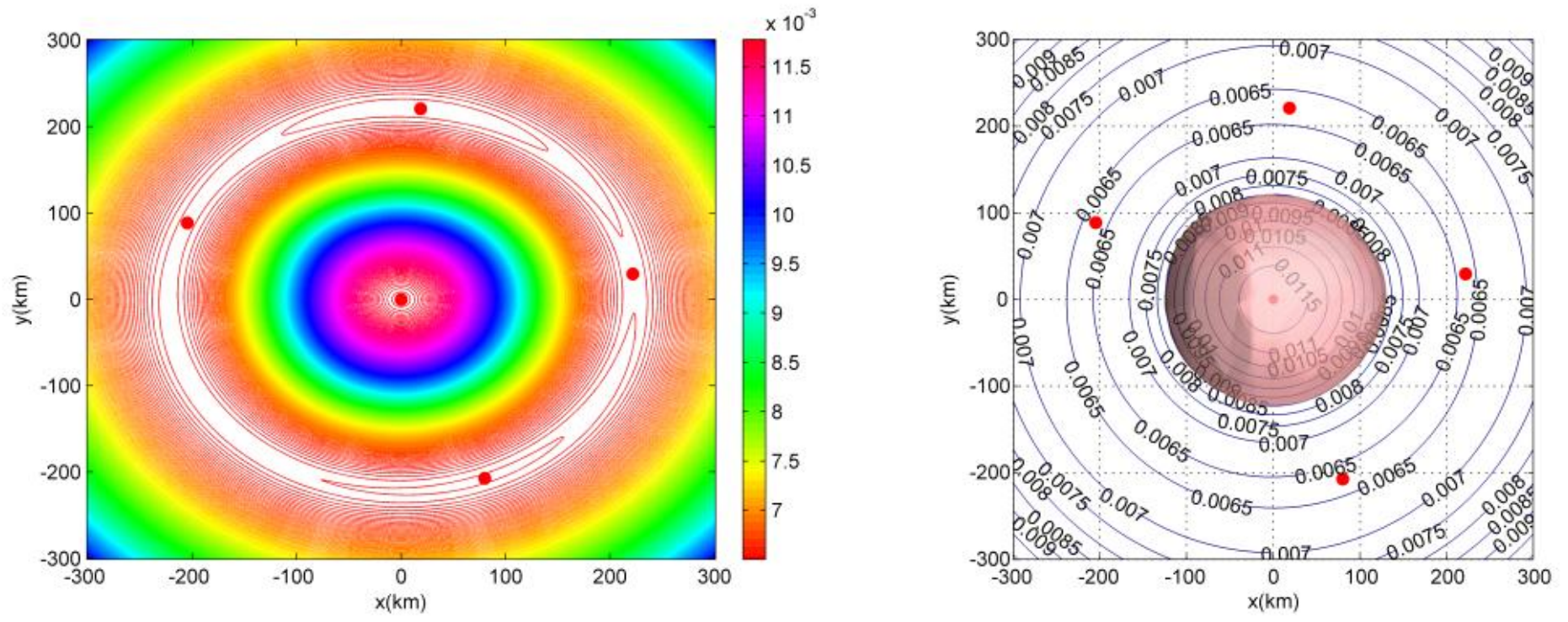

(d)

Figure 4. Zero-velocity curves and equilibrium points with the shape variety of the body. (a) $\eta = 0.8$; (b) $\eta = 0.6$; (b) $\eta = 0.4$; (b) $\eta = 0.2$.

## 4 Simulation of Orbits of the Moonlet

Asteroid (41) Daphne has a moonlet S/2008 (41) 1. The semi-major axis of the moonlet relative to the equatorial inertial coordinate system of (41) Daphne is 443 km [4]. The rotation period of (41) Daphne is 5.98798h [3]. The diameter of the moonlet S/2008 (41) 1 is smaller than 2 km [4]. Using the shape model of (41) Daphne, we calculated the size of (41) Daphne to be 261.42×198.38×181.19 km. Thus the binary system (41) Daphne-S/2008 (41)1 is a large-size-ratio system. To help to understand the motion of the moonlet in the potential of (41) Daphne, we investigated the orbits around (41) Daphne to model the motion of S/2008 (41)1.

Figure 5 shows the simulation of the orbit of moonlet in the potential of 41 Daphne. The time of the orbit is 598.798h. When the asteroid (41) Daphne rotate 100 circles, the simulated orbit is ended. The gravitational field of 41 Daphne is calculated by the polyhedron model generated by observation data. In Figure 5(a), one can see the 3D plot of the orbit which is shown relative to the body-fixed frame of 41 Daphne; while in Figure 5(b), one can see the 3D plot of the orbit which is shown relative to the equatorial inertia system of 41 Daphne. The origin of coordinate of the equatorial inertia system is set as the mass center of 41 Daphne. The x, y, z axes of the equatorial inertia system are paralleled to the x, y, z axes of the body-fixed frame at the initial time. From Figure 5(a) and 5(b), one can see that the trajectories relative to the body-fixed frame and the inertia frame are different. The trajectory relative to the inertia frame has a small change relative to the first circle of the orbit. Which gives a geometric intuition to understand the stability of the orbit for S/2008 (41)1 relative to 41 Daphne.

Figure 5(c) gives the mechanical energy of the orbit while Figure 5(d) gives the Jacobi integral of the orbit. The mechanical energy of the orbit is in the interval of [-491.079, -479.262]J·kg$^{-1}$. The mechanical energy of the orbit has a periodic variation. The Jacobi integral of the orbit is a constant because the system is a conservative system. The value of the Jacobi integral is -4693.725 J·kg$^{-1}$. Our results shows the value of the Jacobi integral is in the interval of [-4693.72499572420, -4693.72499571904]J·kg$^{-1}$.

Chanut et al. [23] studied the stability of 3D plausible orbital stability in the gravitational potential of 216 Kleopatra, which is the primary of a triple asteroid system. They give an example of 3D orbit that keep stable in the gravitational potential of 216 Kleopatra after 1000h. The radius of the orbit is 250km and the eccentricity is 0.2, thus they conclude that the stable orbits exist at a periapsis radius of 250 km and the eccentricity of 0.2. Our work indicates that the stable orbit exist in the gravitational potential of 41 Daphne with the semi-major axis the same as the moonlet's. Considering the irregular shape of 41 Daphne, the effects of the irregular shape to the motion of the particle is significant when the distance between the particle and the mass center of 41 Daphne is not very large. From the figure of effective potential shown in Section 2 and 3, one can see that the value of effective potential varies significantly when the particle is moving near the equilibrium points. Thus if we want the perturbation of the irregular shape of 41 Daphne to the orbit of the particle become small, the particle should be far away from the equilibrium points. Likewise, if we want the sudden change of the mechanical energy of the orbit become unconspicuous, the particle should also be far away from the equilibrium points. For 41 Daphne, the distance between the four outside equilibrium points and the mass center of the body is about 200 km. The distance of the orbit relative to the mass center of 41 Daphne should be more that 250 km.

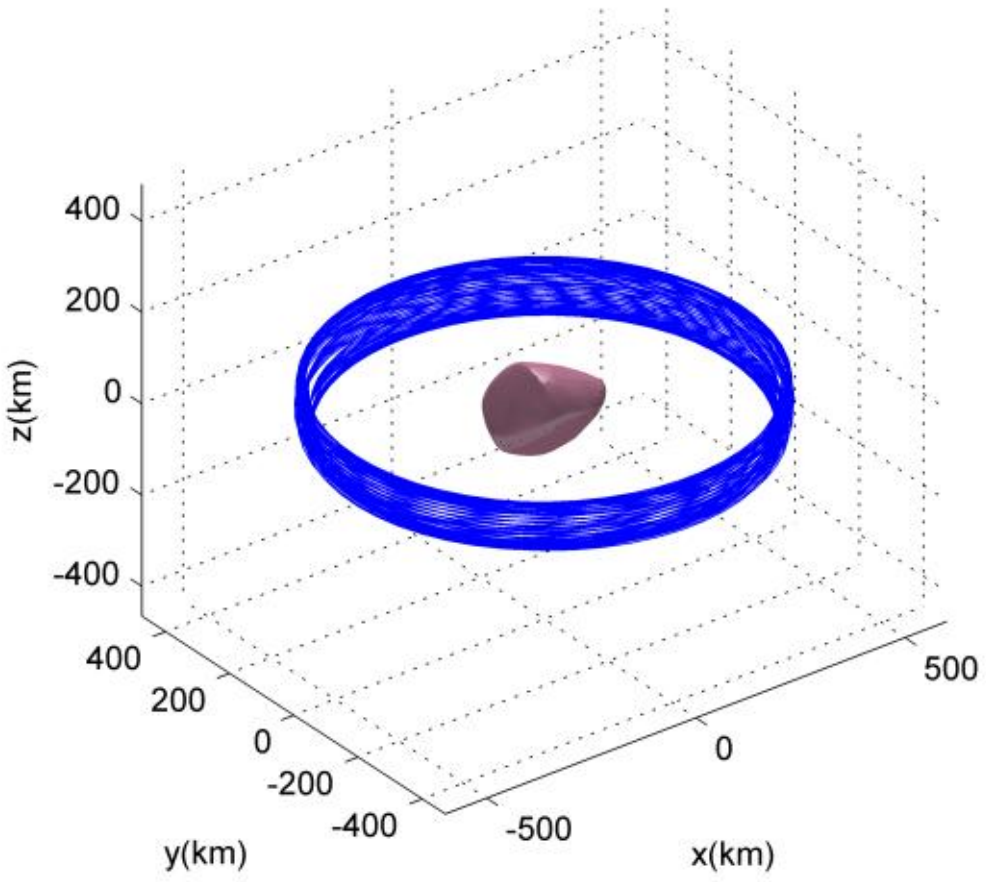

(a)

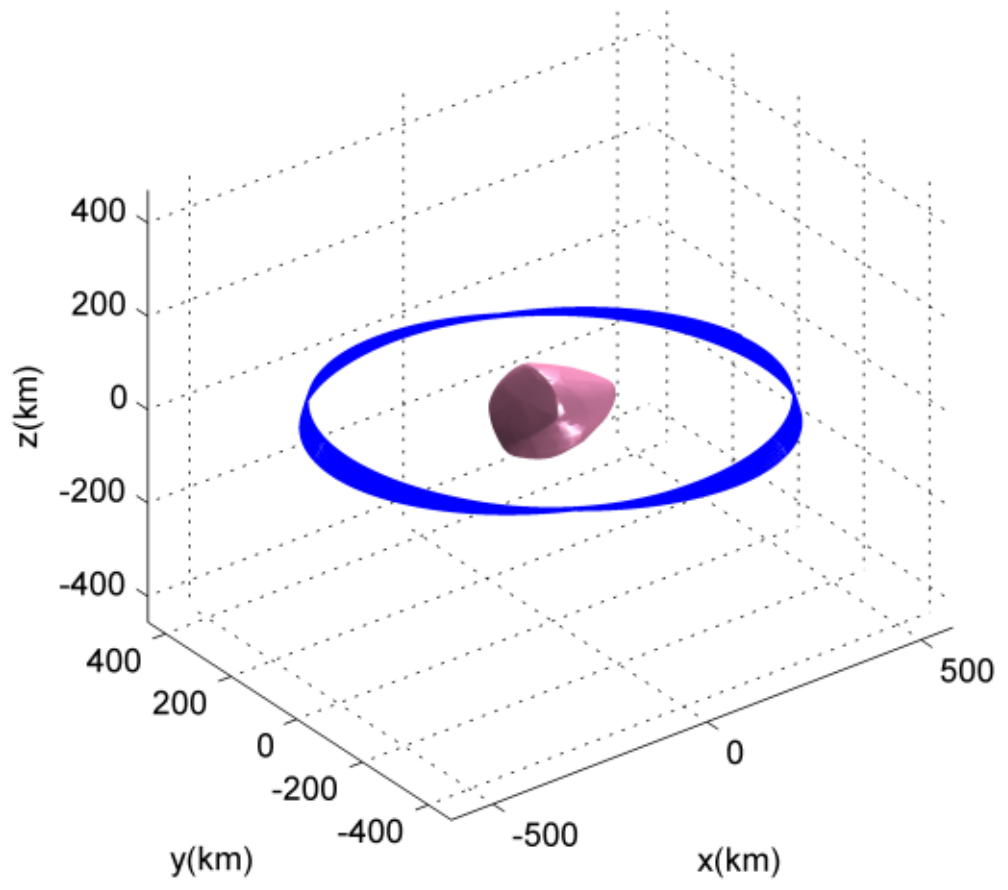

(b)

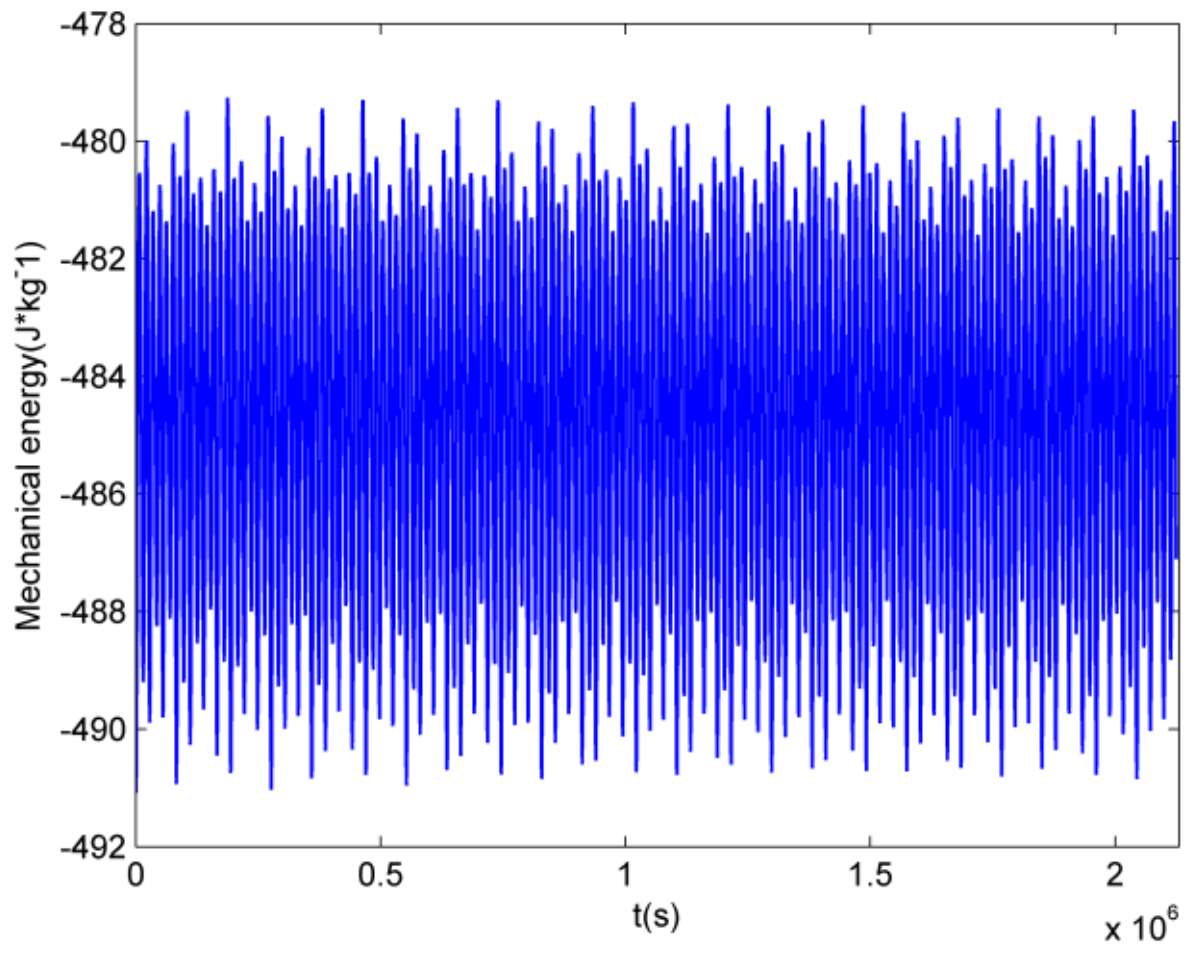

(c)

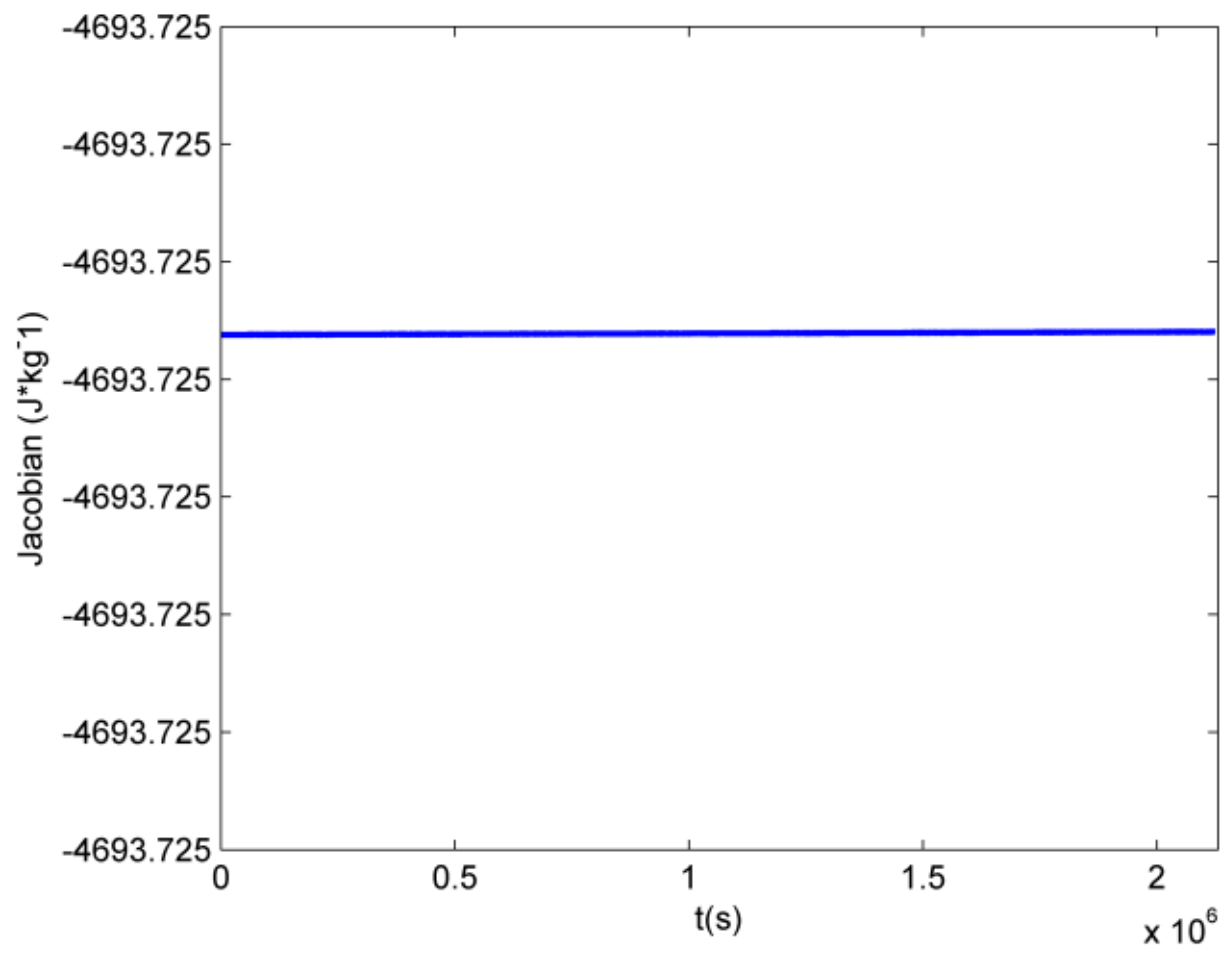

(d)

Figure 5. Simulation of the orbit of moonlet in the potential of 41 Daphne. (a) 3D plot of the orbit in the coordinate system of principal axis of inertia. (b) 3D plot of the orbit in the inertia system. (c) Mechanical energy of the orbit. (d) Jacobi integral of the orbit.

## 5 Conclusion

In this paper, we used the polyhedron model with observational data of the irregular shape to generate the 3D asymmetric convex shape model of asteroid (41) Daphne. In general, for the asteroids which have five equilibrium points, there are two outside equilibrium points near the principal axes of smallest moment of inertia, and the other two outside equilibrium points near the principal axes of intermediate moment of inertia. The inner equilibrium point is near the mass center of the asteroid. More detailed contents of the asteroids with five equilibrium points can be seen in Wang et al. (2014). However, for asteroid (41) Daphne, the positions of two outside equilibrium points associated to the principal axes of smallest moment of inertia have large deviations relative to the axes, the positions of two outside equilibrium points associated to the principal axes of intermediate moment of inertia also have large deviations relative to the axes.

For the asteroid 41 Daphne, the topological cases of the outer equilibrium points E1-E4 are Case 2, Case 5, Case 2, and Case 1. Wang et al. (2014) investigated 23 minor celestial bodies, including 15 asteroids. The results in Wang et al. (2014) showed that the outside equilibrium points belonging to Cases 2 and 5 have a staggered distribution, and the outside equilibrium points belonging to Cases 1 and 2 also have a staggered distribution. Thus the distribution of topological cases of equilibrium points around (41) Daphne is quite different from other minor celestial bodies. Additionally, the outside equilibrium point E4 around (41) Daphne is linearly stable. There is only one linearly stable equilibrium point outside the asteroid (41) Daphne.

The positions, Jacobi integral, zero-velocity surfaces, and eigenvalues vary during the variety of the body from the shape of Daphne to a sphere. The positive definite and non- positive definite as well as the indices of inertia of the Hessian matrix of the equilibrium points remain unchanged. The topological case of the outer equilibrium point E2 changes from Case 5 to Case 1 and its stability changes from unstable to linearly stable.

The orbit for the moonlet in the potential of (41) Daphne has been simulated using the gravitational field computed by the polyhedron model with the observation data of the irregular shape. The mechanical energy of the orbit changes periodically, the Jacobi integral of the orbit keeps conservative.

## Acknowledgements

This project is funded by China Postdoctoral Science Foundation- General Program (No. 2017M610875) and the National Natural Science Foundation of China (No. 11772356).

## Appendix A

Table A1 Hessian matrix of the equilibrium points around 41 Daphne

E1

| | | |
|---|---|---|
| -0.299629390254456E-06 | -0.372702326512410E-07 | -0.442810448250495E-08 |
| -0.372702326512410E-07 | 0.133847271473698E-07 | 0.592684128349602E-09 |
| -0.442810448250493E-08 | 0.592684128349572E-09 | 0.111227472767123E-06 |

E2

| | | |
|---|---|---|
| -0.153488229948157E-07 | -0.232578319597531E-07 | 0.204620208226413E-08 |
| -0.232578319597530E-07 | -0.250256653617802E-06 | 0.114336209257703E-08 |
| 0.204620208226409E-08 | 0.114336209257727E-08 | 0.905882862726529E-07 |

E3

| | | |
|---|---|---|
| -0.253281074912665E-06 | 0.861527786934009E-07 | 0.400512918839414E-08 |
| 0.861527786934010E-07 | -0.205841791790279E-07 | -0.499649803347667E-08 |
| 0.400512918839399E-08 | -0.499649803347674E-08 | 0.988480637517284E-07 |

E4

| | | |
|---|---|---|
| -0.248475425176258E-07 | 0.664802836867783E-07 | 0.157463175261293E-08 |
| 0.664802836867783E-07 | -0.243886117873808E-06 | -0.650357085195925E-09 |
| 0.157463175261286E-08 | -0.650357085195676E-09 | 0.937164700514688E-07 |

E5

| | | |
|---|---|---|
| 0.344108802460873E-06 | 0.120171892849826E-08 | -0.652958818993591E-09 |
| 0.120171892849835E-08 | 0.445791447186455E-06 | -0.479230146376088E-08 |
| -0.652958818993542E-09 | -0.479230146376091E-08 | 0.670576384091542E-06 |

Table A2 Positions and Jacobi integrals of the equilibrium points of the body during the shape variety

$\eta = 0.8$

| Equilibrium Points | x (km) | y (km) | z (km) | Jacobi integral ($1.0\times 10^{-3}km^2s^{-2}$) |
|---|---|---|---|---|
| E1 | 190.839 | 22.9360 | -3.60120 | 4.56987 |
| E2 | 16.6553 | 181.478 | 1.18065 | 4.42397 |
| E3 | -178.487 | 62.0088 | -3.66489 | 4.52802 |
| E4 | 47.8292 | -177.872 | 1.02370 | 4.44930 |
| E5 | -1.13145 | -0.568950 | 0.884128 | 8.09998 |

$\eta = 0.6$

| Equilibrium Points | x (km) | y (km) | z (km) | Jacobi integral ($1.0\times 10^{-3}km^2s^{-2}$) |
|---|---|---|---|---|
| E1 | 200.867 | 24.2153 | -2.97724 | 5.15739 |
| E2 | 16.9461 | 194.558 | 0.912249 | 5.04868 |
| E3 | -188.449 | 67.1784 | -2.96023 | 5.12667 |
| E4 | 51.4458 | -189.931 | 0.778375 | 5.06913 |
| E5 | -0.568544 | -0.283754 | 0.547353 | 9.26147 |

$\eta = 0.4$

| Equilibrium Points | x (km) | y (km) | z (km) | Jacobi integral (1.0× $10^{-3}km^2s^{-2}$) |
|---|---|---|---|---|
| E1 | 211.272 | 25.9293 | -2.33443 | 5.79493 |
| E2 | 17.2545 | 207.686 | 0.663333 | 5.72247 |
| E3 | -197.982 | 74.3353 | -2.26246 | 5.77493 |
| E4 | 58.1572 | -201.235 | 0.521069 | 5.73815 |
| E5 | -0.225267 | -0.103149 | 0.308553 | 10.5071 |

$\eta = 0.2$

| Equilibrium Points | x (km) | y (km) | z (km) | Jacobi integral (1.0× $10^{-3}km^2s^{-2}$) |
|---|---|---|---|---|
| E1 | 221.806 | 29.3465 | -1.68528 | 6.48199 |
| E2 | 18.6169 | 220.772 | 0.433460 | 6.44533 |
| E3 | -204.829 | 88.5246 | -1.54061 | 6.47259 |
| E4 | 80.0730 | -207.319 | 0.210440 | 6.45612 |
| E5 | -0.0428778 | -0.00376719 | 0.152237 | 11.8368 |

Table A3 Eigenvalues of the equilibrium points of the body during the shape variety

$\eta = 0.8$

| Equilibrium Points(× $10^{-3}s^{-1}$) | $\lambda_1$ | $\lambda_2$ | $\lambda_3$ | $\lambda_4$ | $\lambda_5$ | $\lambda_6$ |
|---|---|---|---|---|---|---|
| E1 | 0.328183i | -0.328183i | 0.318324i | -0.318324i | 0.184438 | -0.184438 |
| E2 | 0.300828i | -0.300828i | 0.0568716+ 0.213294i | 0.0568716- 0.213294i | -0.0568716+ 0.213294i | -0.0568716- 0.213294i |
| E3 | 0.318882i | -0.318882i | 0.298832i | -0.298832i | 0.126370 | -0.126370 |
| E4 | 0.304632i | -0.304632i | 0.254473i | -0.254473i | 0.132136i | -0.132136i |
| E5 | 1.00094i | -1.00094i | 0.802277i | -0.802277i | 0.406170i | -0.406170i |

$\eta = 0.6$

| Equilibrium Points(×$10^{-3}s^{-1}$) | $\lambda_1$ | $\lambda_2$ | $\lambda_3$ | $\lambda_4$ | $\lambda_5$ | $\lambda_6$ |
|---|---|---|---|---|---|---|
| E1 | 0.318368i | -0.318368i | 0.310167i | -0.310167i | 0.150149 | -0.150149 |
| E2 | 0.300030i | -0.300030i | 0.236747i | -0.236747i | 0.170147i | -0.170147i |
| E3 | 0.312782i | -0.312782i | 0.296742i | -0.296742i | 0.104266 | -0.104266 |
| E4 | 0.302892i | -0.302892i | 0.270667i | -0.270667i | 0.100064i | -0.100064i |
| E5 | 1.00869i | -1.00869i | 0.787646i | -0.787646i | 0.415533i | -0.415533i |

$\eta = 0.4$

| Equilibrium Points(×$10^{-3}s^{-1}$) | $\lambda_1$ | $\lambda_2$ | $\lambda_3$ | $\lambda_4$ | $\lambda_5$ | $\lambda_6$ |
|---|---|---|---|---|---|---|
| E1 | 0.310233i | -0.310233i | 0.303479i | -0.303479i | 0.115440 | -0.115440 |
| E2 | 0.299235i | -0.299235i | 0.267151i | -0.267151i | 0.118767i | -0.118767i |
| E3 | 0.307650i | -0.307650i | 0.295076i | -0.295076i | 0.0818610 | -0.0818610 |
| E4 | 0.301343i | -0.301343i | 0.280785i | -0.280785i | 0.0732746i | -0.0732746i |
| E5 | 1.01550i | -1.01550i | 0.774645i | -0.774645i | 0.423312i | -0.423312i |

$\eta = 0.2$

| Equilibrium Points(×$10^{-3}s^{-1}$) | $\lambda_1$ | $\lambda_2$ | $\lambda_3$ | $\lambda_4$ | $\lambda_5$ | $\lambda_6$ |
|---|---|---|---|---|---|---|
| E1 | 0.303657i | -0.303657i | 0.297871i | -0.297871i | 0.0769223 | -0.0769223 |
| E2 | 0.298458i | -0.298458i | 0.282279i | -0.282279i | 0.0791089i | -0.0791089i |
| E3 | 0.303367i | -0.303367i | 0.293759i | -0.293759i | 0.0575210 | -0.0575210 |
| E4 | 0.300056i | -0.300056i | 0.287370i | -0.287370i | 0.0490069i | -0.0490069i |
| E5 | 1.02154i | -1.02154i | 0.763046i | -0.763046i | 0.429811i | -0.429811i |

Table A4 The topological classification, stability, and index of inertia of the equilibrium points of the body during the shape variety. LS: linearly stable; U: unstable; P: positive definite; N: non-positive definite; Index of inertia: positive/ negative index of inertia

$\eta = 0.8$

| Equilibrium Points | Topological Case | Stability | $\nabla^2 V$ | Index of Inertia |
|---|---|---|---|---|
| E1 | 2 | U | N | 2/1 |
| E2 | 5 | U | N | 1/2 |
| E3 | 2 | U | N | 2/1 |
| E4 | 1 | LS | N | 1/2 |
| E5 | 1 | LS | P | 3/0 |

$\eta = 0.6$

| Equilibrium Points | Topological Case | Stability | $\nabla^2 V$ | Index of Inertia |
|---|---|---|---|---|
| E1 | 2 | U | N | 2/1 |
| E2 | 1 | LS | N | 1/2 |
| E3 | 2 | U | N | 2/1 |
| E4 | 1 | LS | N | 1/2 |
| E5 | 1 | LS | P | 3/0 |

$\eta = 0.4$

| Equilibrium Points | Topological Case | Stability | $\nabla^2 V$ | Index of Inertia |
|---|---|---|---|---|
| E1 | 2 | U | N | 2/1 |
| E2 | 1 | LS | N | 1/2 |
| E3 | 2 | U | N | 2/1 |
| E4 | 1 | LS | N | 1/2 |
| E5 | 1 | LS | P | 3/0 |

$\eta = 0.2$

| Equilibrium Points | Topological Case | Stability | $\nabla^2 V$ | Index of Inertia |
|---|---|---|---|---|
| E1 | 2 | U | N | 2/1 |
| E2 | 1 | LS | N | 1/2 |
| E3 | 2 | U | N | 2/1 |
| E4 | 1 | LS | N | 1/2 |
| E5 | 1 | LS | P | 3/0 |